\documentstyle[twocolumn,seceq]{jpsj}

\def\runtitle{Submillimeter Wave ESR Study of Spin Gap Excitations in CuGeO$_3$}
\def\runauthor{Hiroyuki {\sc Nojiri}, Hitoshi {\sc Ohta},Susumu {\sc Okubo}, Osamu {\sc Fujita}, Jun {\sc Akimitsu}, and
Mitsuhiro {\sc Motokawa}}

\title
{
Submillimeter Wave ESR Study of Spin Gap Excitations in CuGeO$_3$
}

\author
{ 
Hiroyuki {\sc Nojiri}\, Hitoshi {\sc Ohta}$^{1,}$\, 
Susumu {\sc Okubo}$^{2,}$\,  Osamu {\sc Fujita}$^{3,}$\,  Jun {\sc Akimitsu}$^{3,}$\ and
Mitsuhiro {\sc Motokawa}\
}
\inst
{
Institute for Materials Research, Tohoku University, Katahira 2-1-1, Sendai 980-8577, Japan\\
$^1$Department of Physics, Kobe University, Rokkodai 1-1, Kobe 657, Japan\\
$^2$The Graduate School of Science and Technology, Kobe University, Rokkodai 1-1, Kobe 657,
Japan\\
$^3$Department of Physics, Aoyama-Gakuin University, Chitosedai 6-16-1, Setagaya-ku, Tokyo
157-8572, Japan\\ }

\recdate
{
 ? 
}

\abst
{
Transitions between the ground singlet state to the excited triplet state has been observed in
CuGeO\(_3\) by means of submillimeter wave electron spin resonance. The strong absorption
intensity shows the break down of the selection rule. The energy gap at zero field is evaluated
to be 570 GHz(2.36 meV) and this value is nearly identical to the gap at the zone center
observed by inelastic neutron scattering. The absorption intensity shows strong field
orientation dependence but shows no significant dependence on magnetic field intensity. These
features have been explained by considering the existence of Dzyaloshinsky-Moriya (DM)
antisymmetric exchange interaction. The doping effect on this singlet-triplet excitation has
been also studied. A drastic broadening of the absorption line is observed by the doping of only
0.5 \% of Si.
}

\kword
{
CuGeO\(_3\), submillimeter wave ESR, magnetic excitation, spin gap, high magnetic field
}

\begin{document}
\sloppy
\maketitle

\section{Introduction}
CuGeO\(_3\) has attracted considerable attentions as the first inorganic material which shows 
a spin-Peierls transition.\cite{rf:1} 
A large number of experimental and theoreticalstudies have been made in these few years on this compound and the study of
magnetic excitations has been one of the most interesting subjects.
 Among many different kinds of experimental methods to study magnetic excitations such as inelastic neutron
scattering\cite{rf:2,rf:3,rf:4,rf:5,rf:6}, Raman scattering or far-infrared spectroscopy\cite{rf:7}, ESR is considered to be a
unique tool for the following characteristics:(1) high sensitivity, (2) high energy resolution and (3)possibility of high field
measurements by using a pulsed magnetic field. 
The third feature of ESR enables us to investigate magnetic excitations in all three fundamental magnetic phases such as uniform
phase(U-phase), dimerized spin-Peierls phase(SP-phase) and so called magnetic phase(M-phase). 
As the matter of fact, many ESR
investigations have been reported on CuGeO\(_3\) and different aspects of magnetic excitations have been
discussed.\cite{rf:8,rf:9,rf:10,rf:11,rf:12,rf:13}  However, only a few direct observations were made for the spin gap
excitation.\cite{rf:8,rf:14,rf:15}
 The one of the reasons is that the transition between the ground singlet state and the excited triplet state(we call this
transition as the singlet-triplet transition hereafter) is forbidden as ESR in principle because the total spin is conserved
in magnetic dipole transition.
 Moreover, the energy gap is as high as a few meV and it requires a special ESR equipment which covers the high frequency
submillimeter wave. 
For these reasons, it was considered that the observation of the singlet-triplet transition was very difficult. 

An observation
of the singlet-triplet transition in CuGeO\(_3\) by ESR was first reported by Brill {\it et al}. under magnetic fields parallel
to the  {\mib a}-axis.\cite{rf:8}
 The value of the energy gap was reported as  {\it E}\(_{g1}\)=63.77 K(5.495 meV) and it was explained as the transition at the 
{\mib q}={\mib 0}. 
The field orientation dependence of this singlet-triplet transition were also reported by a different
group.\cite{rf:14}

	Since the early stage of researches on CuGeO\(_3\), we have performed a systematic ESR investigation on three fundamental
magnetic phases by using our submillimeter wave ESR equipment which covers both high magnetic field region and wide frequency
range.
 In the course of this study, we have found a new strong singlet-triplet transition with an energy gap of {\it
E}\(_{g2}\)=2.36 meV.\cite{rf:15}
 This gap energy {\it E}\(_{g2}\) is different from that observed by Brill {\it et al}.\cite{rf:8} and it is close
to the spin gap energy at zone center [0, 1, l/2] which was observed by inelastic neutron scattering(INS).
 The absorption intensity of newly found singlet-triplet transition is very strong and this fact is considered to be strange
because the singlet-triplet transition is usually forbidden as explained above.
 Moreover, the absorption intensity shows a strong field orientation dependence.
 These features show that the spin-Hamiltonian of the system is much deviated from the simple Heisenberg type and show that a
non-secular term such as Dzyaloshinsky-Moriya interaction should be included into that Hamiltonian.

	Before proceeding to the experimental results, we describe briefly about the dispersion of magnetic excitation in CuGeO\(_3\)
observed by INS experiments in dimerized SP phase.\cite{rf:2,rf:3,rf:4,rf:5,rf:6}
 A very steep dispersion is observed along the magnetic chain of the  {\mib c}-axis.
 In addition to this, a two-dimensional modulation of the dispersion as shown in Fig. 1 is observed for the non-negligible
interchain exchange coupling {\it J}\(_b\) along the {\mib b}-axis.
 For the simplicity, we use a notation  {\mib q}=({\it k}, {\it l}) which is the wave vector associated with the two dimensional
reciprocal space spanned by {\mib b*} and  {\mib c*}.
 In this notation the energy gap {\it E}\(_{g2}\)=2.36 meV is related to the gap at  {\mib q}=(0,$ \pi $) or {\mib  {\mib q}}=($
\pi $,0) and the energy gap {\it E}\(_{g1}\)=5.50 meV is related to that at  {\mib q}=(0,0) or  {\mib q}=($ \pi $, $ \pi $).
This two-dimensional dispersion relation indicates that the newly observed ESR peak is related to the energy gap at
the reciprocal lattice point where the wave vector is not zero.
 This fact implies that the magnetic excitation at  {\mib q}=(0,$ \pi $) or  {\mib q}=($ \pi $,0) are folded towards  {\mib
q}=(0,0) in the dimerized SP phase. Another important feature of the magnetic excitation of CuGeO\(_3\) is that the existence
of the double gap structure where the first excited triplet branch is separated by the second gap from the two magnon spin wave
continuum.\cite{rf:6}
 This separation enables us to apply a simplified singlet-triplet energy scheme to describe the magnetic excitation in dimerized
SP phase of CuGeO\(_3\).

   In the following, we report the results on the newly found singlet-triplet transition of {\it E}\(_{g2}\) in
CuGeO\(_3\).  The temperature, field intensity and field orientation dependencies of this excitation will be shown.
 The origin of the breaking of the selection rule will be discussed considering the experimentally observed features.
 Finally the effect of doping will be also shown.

\section{Experimental}
Submillimeter wave ESR measurements have been performed up to 800 GHz in pulsed magnetic fields up to 30 T. A far-infrared
laser, backward traveling wave tubes and Gunn oscillators have been employed as the radiation source. An InSb is used as a
detector. Both the solenoid and the split type pulsed magnets have been used for the field generation. Single crystal samples
were synthesized by the floating zone technique. The concentration of Si is determined by EPMA method.

\section{Results}
	
\subsection{Singlet-triplet transition}
	Figure 2 shows examples of ESR spectra measured at 1.7 K in three different magnetic field orientations.  All the measurements
were performed in the Faraday configuration where the propagation vector {\mib k} of the radiation is parallel the external
magnetic field.

The absorption lines denoted by TR is the transition within the excited triplet state and the signal is caused by the thermally
activated spins across the energy gap.
 Another transitions marked by M is the ESR mode in the M phase where the energy gap is collapsed by strong magnetic
fields.\cite{rf:9, rf:11}  The sharp absorption lines indicated by P are the signal of DPPH used as the field calibration maker.
 The resonance fields of TR or M absorption lines increase as frequency is increased.
 Since these two kinds of ESR modes have been already reported in separated papers, we do not explain them further.
 There is an another type of resonance peak marked by D with broad line width.
 For the signal D, the resonance field decreases as the frequency increases up to 570 GHz and above this frequency the
resonance field increases as the frequency is increased.
 The frequency field diagram of this transition is shown in Fig. 3 for all three magnetic field orientations.
 This figure shows that signal D is associated with the zero field energy gap of {\it E}\(_{g2}\)=570 GHz(2.36 meV).
 A large zero field gap shown in Fig. 3 is not expected for a paramagnetic resonance of {\it S}=1/2 Cu$^{2+}$ ions.
 Moreover this energy coincide with the value of spin gap obtained by neutron scattering at {\mib q}=($ \pi $, 0) or {\mib
q}=(0, $ \pi $) as mentioned before.
 These facts indicate that the signal D is the direct transition between the ground state and the excited triplet state as shown
schematically in the inset of Fig. 3.

 To confirm that signal D is the singlet-triplet transition, we have also measured the temperature dependence of the ESR spectra
at 342.0 GHz as shown in Fig. 4.
 The intensity of the signal D located around 8 T increases as the temperature is decreased.
This behavior clearly indicates that the signal D is the transition from the ground state and not a transition among the
excited states. The temperature dependence of signal TR appears around 12 T is completely opposite to that of signal D.
 In the case of signal TR, the absorption intensity increases as the temperature is increased showing obviously that this signal
is related to the thermally excited states.
	Table 1 shows the list of zero field energy gap value and {\it g}-value of each triplet blanch.
 The obtained {\it g}-values are consistent with the values reported in the references.\cite{rf:8,rf:9,rf:10,rf:11}
 To examine the existence of the anisotropic exchange interaction, we carefully check if a zero field energy splitting exists
between {\it S}\(_{z}\)=1 and {\it S}\(_{z}\) =-1 branches.
 However, the energy gap of observed six branches are in the range of 570$\pm$2 GHz and no significant zero field splitting is
found. The {\it S}\(_{z}\)=0 branch has not been because the frequency of this branch is field independent and thus it cannot be
detected by our magnetic field sweep ESR.

	Another very important feature of this transition is that the intensity is anisotropic as shown in Fig. 2.
 The absorption intensity is strongest for {\mib B}$\parallel${\mib c} and weakest for {\mib B}$\parallel${\mib b}.
 Since there are two inequivalent Cu$^{2+}$ sites in CuGeO$_3$, a staggered moment can be induced by external magnetic field.
 However, two sites are equivalent with respect of the magnetic field orientation in the case of {\mib B}$\parallel${\mib c}.
 Hence, the fact that the absorption intensity is strongest for {\mib B}$\parallel${\mib c} and a staggered field mechanism as
the candidate of the breaking of the selection rule are not compatible.
 This point will be discussed later.

\subsection{Field dependence}

	To study the possible softening of the lattice dimerization around the critical field {\it H}$_c$ between SP and M phase,
 we check the non-linear behavior of the triplet branch. If softening occurs, the energy gap depends on field intensity and
then the field dependence of the resonance frequency becomes non-liner.
 We have tried the fitting of the observed {\it S}\(_{z}\)=-1 branch for {\mib B}$\parallel${\mib c} assuming a power law
field dependence of energy gap as 

\begin{equation}
h\nu=\Delta{_0}\delta H{^{\alpha}}-g\mu{_B} S{_z}H
\end{equation}

where $\nu$ is the frequency of ESR, $\Delta{_0}$ is the energy gap at zero field, $\delta$H is a normalized magnetic field
as
$\delta$H =({\it H}$_c$-{\it H})/{\it H}$_c$,
 {\it H}$_c$ is the critical field between SP and M phases and {\it H} is the external magnetic field. The fitting gives a
very small value of $\alpha$=2.5$\times$10$^{-4}$ when 0$\le${\it B}$\le$12.7 T.
 The fitting with different magnetic field range such as 10$\le${\it B}$\le$12.7 T shows no significant change of
 $\alpha$. This fact show that there is no field dependence of energy gap at least when $\delta${\it H}$\geq$0.01 within the
experimental error and indicates that the transition between SP phase and M phase is the first order transition.

 This first order nature of the
transition is also shown in the hysteresis of the signal D as shown in Fig. 5.
 At 195.8 GHz, the absorption intensity as well
as resonance field changes between increasing and decreasing fields.
 This change is consistent with the hysteresis of the
magnetization process. At slightly lower frequency of 190 GHz, the direct transition suddenly disappears becuase the
spin-Peierls energy gap collapeses in the M phase.
 Considering above mentioned field dependence of the singlet-triplet
transition, we conclude that no significant softening of spin gap is induced by magnetic fields in the field range of
0.01$\geq$$\delta${\it H}.

 Another important point is that the down going branch of the excited triplet does not show a level crossing with the
ground singlet state at {\it H}$_c$ as shown in Fig. 3.
 It means that the ground state of the system changes from non-magnetic to magnetic states at {\it H}$_c$
with a finite energy gap.
 This behavior is different from that of a simple isolated dimer spin system where the zero
field energy gap {\it E}\(_{g}\) and the critical field {\it H}$_c$ holds a simple relation of {\it E}\(_{g}\)={\it
g}$\mu$$_B${\it H}$_c$.
 As is well known, the spin-Peierls transition
costs the lattice deformation energy {\it E}$_d$.
 Considering this loss, we may rewrite the above relation to {\it E}\(_{g}\)-{\it E}$_d$= {\it g}$\mu$$_B${\it
H}$_c$ for the spin-Peierls system.
 This relation expresses that the critical field is decreased by the lattice dimerization energy loss for spin-Peierls system.
 This relation is not precise because an incommensurate lattice dimerization exists even above {\it H}$_c$.
 However, we can consider {\it E}$_d$ as a measure of lattice dimerization energy of the system at SP phase.
 If we assume that {\it E}$_d$ is identical with the energy difference between {\it S}\(_{z}\)=-1 branch and the ground state
at {\it H}$_c$, we obtained {\it E}$_d$=193 GHz.
 By using this {\it E}$_d$, we found a very simple relation between two energy gap {\it E}\(_{g1}\) and {\it E}\(_{g2}\) as {\it
E}\(_{g1}\)=2{\it E}\(_{g2}\)+{\it E}$_d$.
 This point will be discussed later.

 \subsection{Impurity effect} 
	The effect of impurity doping for the singlet-triplet transition have been examined using single crystals in which Ge site
 is partly substituted by Si. The spectra for different Si concentrations are shown in Fig. 6. 
The N{\'{e}}el temperatures and the
spin-Peierls transition temperatures are 2.5 K and 11.5 K for {\it x}=0.01 and 0.9 K and 13.5 K for {\it x}=0.005 samples,
respectively.
 No additional extrinsic signals are observed for both samples. This fact shows the high quality of the single crystals used for
the present experiments.

 The drastic change of the spectrum is found by Si doping and no absorption is observed for {\it x}=0.01.
 In the case of {\it x}=0.005, only a very weak and broad resonance is observed.
 The strong suppression of the singlet-triplet transition shows that the density of state of the first excited triplet state is
very sensitive to the impurity doping.
 In the coexisting phase of Si doped sample, a well defined antiferromagnetic resonance(AFMR) has been
observed.\cite{rf:16}
 This AFMR is related to a spin wave excitation at {\mib q}={\mib 0}. Thus the suppression of the singlet-triplet transition
shows that the spectral weight shifts drastically from the triplet branch to the AFMR mode by a very small amount of Si
doping.

\section{Discussion}
\subsection{Dzyaloshinsky-Moriya interaction} 
	As mentioned previously, the intensity of a singlet-triplet transition is supposed to be very weak.
 However, present results show that the absorption intensity of this forbidden transition is strong in CuGeO\(_3\). As is well
known, a non-secular term should be included into spin-Hamiltonian to cause a singlet-triplet transition. Possible candidates
for such a non-secular term are the following:(1) DM interaction, (2) Staggered moments caused by the alternation of
{\it g}-tensors (3) Anisotropic exchange(AE) interaction. The staggered field effect can be excluded for following reasons. As
was shown by Shiba and Sakai for a Haldane gap system NENP, two distinct features are expected for this
mechanism.\cite{rf:17,rf:18} Firstly, since the staggered magnetic moments are induced by uniform magnetic fields, the
intensity of the singlet-triplet transition should increase as the field is increased. Secondly, the induced moments are
coupled with radiation for Voight configuration where {\mib k}$\perp${\mib B}. In the present case, no strong field dependence
of absorption intensity is observed as shown in Fig. 2. To evaluate the relationship between the intensity of singlet-triplet
excitation and the direction of a magnetic field or propagation vector, we have also performed the experiments in Voight
configuration as shown in Fig. 7. The results for both Faraday and Voight configurations are summarized in table 2. As shown in
table 2, the transition is observed for both Faraday and Voight configuration. For example, the transition is very strong for
{\mib B}$\parallel${\mib k}$\parallel${\mib c}. In this configuration, as mentioned in previous section, the staggered moment
is not expected. Thus we conclude that the staggered field mechanism cannot be applied to the present case.

 Next we consider the possible breaking of selection rule due to DM interaction.
 As is well known, both DM interaction and AE interaction are related to the spin-orbit coupling $\lambda$. It should be noted
that DM interaction causes a stronger effect than AE interaction because the former term is the first order correction of
spin-orbit coupling $\lambda$ while the latter is the second order correction. Thus if DM interaction exists in CuGeO\(_3\),
this will be a leading term as the origin of the experimentally observed singlet-triplet transition. Moreover, the fact that the
zero field splitting of energy gap due to AE was not found shows that AE may be small in the present system. The existence of
the DM interaction for U phase was first proposed by Yamada {\it et al}. by considering the characteristic behavior of the line
width of EPR.\cite{rf:13}  Although the known crystal structure at that time was not compatible with the existence of DM
interaction, a reexamination of crystal structure was made by X-ray diffraction and it was shown that the new crystal structure
allows the existence of DM interaction.\cite{rf:19} According to the proposal of Yamada {\it et al}, ${\mib D}$-vectors are
parallel to one another along the ${\mib a}$-axis and they are alternated along the $\mib b$-axis with an angle of about 100
degree pointing to one of the principal axes of CuO$_6$ octahedron as depicted in the inset of Fig. 7. This complicated
arrangement is caused by the existence of inequivalent Cu sites. Next we consider the direction of ${\mib D}$-vectors in
dimerized SP phase. Considering the lattice deformation in SP phase determined by neutron scattering, we can expect that the
direction of ${\mib D}$-vectors are not changed by the dimerization.\cite{rf:20} Thus the existence of ${\mib D}$-vectors are
justified for SP phase.

   Recently Kokado and Suzuki have made a theoretical calculation of ESR intensity considering DM interaction and proposed 
the following features;(1) the absorption intensity shows no strong field dependence, (2) The transition can be observed in the
Faraday configuration (3) The intensity is strongest when {\mib D}$\perp${\mib B} and weakest when ${\mib D}$$\parallel${\mib
B}.\cite{rf:21}
 As mentioned before, the theoretically
expected two features (1) and (2) are clearly consistent with our experiments. We now discuss the third point considering the
direction of the proposed ${\mib D}$-vectors. For simplicity, we discuss the results in the Faraday configuration. In the
experiment, the intensity is strongest for ${\mib B}$$\parallel${\mib
c} and weakest for ${\mib B}$$\parallel${\mib
b}. In the ${\mib B}$$\parallel${\mib
c} configuration the condition {\mib
D}$\perp${\mib B} is satisfied because the ${\mib D}$ vectors are in the ${\mib {ab}}$-plane. Thus we can expect the strong
singlet-triplet transition for this configuration. If we compare two
configurations ${\mib B}$$\parallel${\mib a} and ${\mib B}$$\parallel${\mib b}, the component of the ${\mib D}$-vector in the
plane normal to the magnetic field is larger for ${\mib B}$$\parallel${\mib a} and smaller for ${\mib B}$$\parallel${\mib b}.
Thus we can expect that the intensity is stronger for ${\mib B}$$\parallel${\mib a} and weaker for ${\mib B}$$\parallel${\mib
b}. Although the intensity of the transition may depend on the details of the spin-Hamiltonian used for the calculation,
above mentioned fundamental and qualitative features shows a good agreement with the present experimental results. Hence we can
conclude that the strong singlet-triplet transition is caused by the DM interaction. Finally we mention that we can not exclude
the possible small contribution of the AE interaction in the present case. However, as discussed above, we find that the AE is
small for CuGeO$_3$ and thus we conclude that the DM interaction is the leading term.  
\subsection{Zone folding}
  As discussed above, we can expect a finite transition matrix element between the ground singlet state and the excited triplet 
states by including the DM interaction in to the Hamiltonian. However, the gap observed in the present work is related the gap
at the non zero wave vector such as ${\mib q}$=($\pi$,0) or ${\mib q}$=(0,$\pi$). Since only the mode at {\mib q}={\mib 0} can
be excited by electromagnetic wave, we cannot relate our spin gap excitation directly to the result of INS experiments. A
possible explanation to overcome this difficulty is to consider the folding of the magnetic excitation branch by DM
interaction or by lattice dimerization. If the magnetic excitation at ${\mib q}$=($\pi$,0) or ${\mib q}$=(0,$\pi$) is folded to
${\mib q}$=(0,0) in the dimerized SP phase, we can observe this folded magnetic excitation by ESR. Recently J. E. Lorenzo {\it
et al}. found the folding of magnetic excitation in CuGeO\(_3\) by neutron diffraction experiments.\cite{rf:22} They found that
two triplet peaks exist at
${\mib q}$=($\pi$,3/2$\pi$) and ${\mib q}$=(0,3/2$\pi$). Although it is difficult to observe the magnetic excitation at ${\mib
q}$=(0,0) directly by neutron, this observation shows clearly that the zone folding exist in the SP phase of CuGeO$_3$. This
fact supports our explanation that the zone folding makes it possible to observe two spin gap excitations at {\mib q}={\mib 0}
by means of ESR.  
  
\subsection{Relation of two energy gaps} 
As mentioned in the previous section, we have found a simple relation {\it E}\(_{g1}\)=2{\it E}\(_{g2}\)+ Ed experimentally
between  two energy gaps {\it E}\(_{g1}\)=1329 GHz and {\it E}\(_{g2}\)=570 GHz. Figure 8 shows a schematic energy diagram for
two energy gaps. As the values of gaps are determined by ESR, the error is less than a few GHz.\cite{rf:8,rf:14} In the present
experiments, the disappearance of the singlet-triplet transition occurs between 195.8 and 190 GHz. Then we can evaluate E$_d$ as
the average of these two frequencies and it turned out to be E$_d$ =193 GHz. If we calculate 2{\it E}\(_{g2}\)+ E$_d$ by using
our experimental values, we obtain 2{\it E}\(_{g2}\)+ E$_d$=1333 GHz. This is identical to {\it E}\(_{g1}\)=1329 GHz
obtained by Brill {\it et al}. within the experimental errors.\cite{rf:8} Moreover, it is surprising that up-going branch of
the lower triplet and the down-going branch of the higher triplet crosses at the critical field {\it H}$_c$ as showun
schematically in Fig. 8. Although we have no good explanation for this unexpected relation between and {\it E}\(_{g1}\) and {\it
E}\(_{g2}\), we speculate that this simple relation between two spin gaps is not accidental. In the theoretical investigations
of magnetic excitations in CuGeO\(_3\), the dispersion along the
$\mib b$-axis has been treated so far as a modification for the one-dimensional system. However, the present results show that
two dimensional nature of CuGeO\(_3\) is essential and should be considered more seriously.

	Finally we would propose a possible candidate to explain the broad line width.
 As is shown in Fig. 2, the line width of a direct transition is broader than that of TR transition. These two signals are both
related to the triplet state. However, the coupling to the lattice may be different between two cases. The triplet state caused
by electromagnetic radiation may not be followed by releasing of a lattice dimerization because the frequency of the radiation
is much faster that that of phonon which cause the deformation of the lattice. On he other hand, the thermally activated
transitions is coupled with the lattice relaxation through phonon. In another word, former process can be adiabatic with
respect to the lattice. If this speculation is correct, the life time of the triplet excited electromagnetically is expected to
be short and as a result the line width becomes broad.

	To summarize, a strong transition from the ground singlet to the excited triplet has been observed for
 the first time by ESR. The energy gap at zero field is isotropic and it turned out to be 570$\pm$2GHz. The characteristic
features of this transition can be understood by considering DM interaction. The observation of the spin gap transition related
to the gap other than {\mib q}={\mib 0} indicate the folding of the magnetic excitations in the SP phase.

\section*{Acknowledgements}
	We would like to express our gratitude thanks to Prof. I. Harada, Prof. I. Yamada, Prof. N. Suzuki, Dr. S. Kokado
 and Prof. J-P Boucher for valuable discussions and suggestions. This work was supported by
Grant-in-Aid of Ministry of Education, Science, Sports and Culture. 

References

\section{Figure captions}

Fig.1 A schematic dispersion curve of the lowest triplet branch in the {\mib b}$^*${\mib c}$^*$-plane. 
Two different energy gap {\it E}\(_{g1}\)=5.50 meV and {\it E}\(_{g2}\)=2.36 meV exist for the 
interchain coupling {\mib J}$_b$

Fig. 2 Examples of ESR spectra at 1.7 K. TR and M denote the transition among the excited triplet states and
ESR in the M phase, respectively. Broad resonance lines D are the singlet-triplet transition. Mark P indicate
the signals of DPPH used for the field calibration. Number attached to each spectrum is the frequency in the unit of GHz.
The baseline of each spectrum is shifted vertically for convenience.

Fig. 3 The frequency field diagram for the signal D. The up-going branch and down-going branch
 are related to  {\it S}\(_{z}\)=1 branch and
{\it S}\(_{z}\)=-1 branch of the excited triplet state, respectively.

Fig. 4 Temperature dependence of the ESR spectrum at $\nu$=342.0 GHz. The signal located around 8 T is the singlet-triplet
transition D and the signal TR is also observed around 12 T. Arrows indicate the signals of DPPH used for the field
calibration. The baseline of each spectrum is vertically shifted for convenience.

Fig. 5 Hysteres of the singlet-triplet transition observed around the phase boundary between the SP and M phases. The
upper trace and lower trace correspond to the increasing field and the decreasing field processes, respectively. The broad
absorption lines marked as D are the singlet-triplet transition and the sharp absorption lines TR beside the DPPH signals are
the transition with in the excited triplet states. The baseline of each spectrum is vertically shifted.

Fig. 6 The change of the singlet-triplet transition spectrum by Si doping. The {\it x} denotes the concentration of Si. The
vertical scale is magnified ten times for lower two traces. The baseline of each spectrum is vertically shifted.

Fig. 7 An example of field orientation dependence of the singlet-triplet transition. The measurement was made in Voight
configuration where  ${\mib B}$$\perp${\mib k}. The number indicate the angle from the {\mib a}-axis in {\mib {ab}}-plane. The
inset shows the schematic view of the direction of DM vectors {\mib D} which are denoted by thick arrows. Two inequivalent
CuO$_6$ octahedron sites are marked by $\sharp$1 and  $\sharp$2. Shadowed and open circles show the Copper and Oxygen,
respectively. The baseline of each spectrum is vertically shifted

Fig. 8 A schematic energy diagrum of two set of triplets with zero field gaps of {\it E}\(_{g1}\) and  {\it E}\(_{g2}\). The
{\it g}-value is normalized to 2. The dashed line shows the critical field  {\it H}$_c$ between SP phase and M phase and it is
also normalized by using experimentally obtained {\it g}-values listed in table 1.
\begin{table}
\caption{The list of the zero field energy gap and {\it g}-value}
\label{table:1}
\begin{tabular}{@{\hspace{\tabcolsep}\extracolsep{\fill}}cccc} \hline
Field Orientation & {\it S}\(_{z}\)  & Energy Gap(GHz) & {\it g}-value \\ \hline
${\mib B}$$\parallel${\mib a}	& 1	& 570.(0)& 2.15\\
${\mib B}$$\parallel${\mib a}	& -1	& 571(7)& 2.16\\
${\mib B}$$\parallel${\mib b}	& 1	& 568.(2)& 2.23\\ 
${\mib B}$$\parallel${\mib b}	& -1	& 569.(3)& 2.26\\
${\mib B}$$\parallel${\mib c}	& 1	& 571.(9)& 2.05\\
${\mib B}$$\parallel${\mib c}	& -1	& 570.(4)& 2.07\\ \hline
\end{tabular}
\end{table}
\begin{table}
\caption{The list of intensity of the singlet-triplet transition for different magnetic field orientation and for
different direction of a propagation vector. The diagonal components are related to the Faraday configuration and others are
related to the Voight configuration. S, M, W show that the intensity of the singlet-triplet transition is strong, moderate and
weak, respectively}
\label{table:2}
\begin{tabular}{@{\hspace{\tabcolsep}\extracolsep{\fill}}cccc} \hline
Polarization & ${\mib B}$$\parallel${\mib a} & ${\mib B}$$\parallel${\mib b} & ${\mib B}$$\parallel${\mib c} \\ \hline
${\mib k}$$\parallel${\mib a}	& M	& M& M\\
${\mib k}$$\parallel${\mib b}	& M	& W& S\\
${\mib k}$$\parallel${\mib c}	& S	& W& S\\ \hline
\end{tabular}
\end{table}

\end{document}